\documentclass[useAMS,natbib]{mn2e}
\usepackage{amsmath}
\usepackage{graphicx}
\def\sgn{\mathop{\rm sgn}\nolimits} 
\title[Universal properties of dark matter haloes and self similarity.]{Dark matter haloes and self similarity}
\author[Alard, C.]{Alard, C. \\
Institut d'Astrophysique de Paris, 98bis boulevard Arago, 75014 Paris, France \\
email:alard@iap.fr}
\begin{document}
\date{}
\maketitle
\begin{abstract}
 This papers explores the self similar solutions of the Vlasov-Poisson system and their relation to the gravitational
 collapse of dynamically cold systems. Analytic solutions are derived for power law potentials in one dimension, and 
 extensions of these solutions in three dimensions are proposed. Next the self similarity of the collapse of cold dynamical
 systems is investigated numerically. The fold system in phase space is consistent with analytic self similar solutions, 
 the solutions present all the proper self-similar scaling. An additional point is the appearance of an $x^{-\frac{1}{2}}$ 
 law at the center of the system for initial conditions with power law index larger than $-\frac{1}{2}$ (the Binney conjecture).
 It is found that the first appearance of the $x^{-\frac{1}{2}}$ law corresponds to the formation of a singularity very close to
 the center. Finally the general properties of self similar multi dimensional solutions near equilibrium are investigated. Smooth
 and continuous self similar solutions have power law behavior at equilibrium. However cold initial conditions result in discontinuous
 phase space solutions, and the smoothed phase space density looses its auto similar properties. This problem is easily solved by observing
 that the probability distribution of the phase space density $P$ is identical except for scaling parameters to the probability
 distribution of the smoothed phase space density $P_S$. As a consequence $P_S$ inherits the self similar properties of $P$.
 This particular property is at the origin of the universal power law observed in numerical simulation for $\frac{\rho}{\sigma^3}$. 
 The self similar properties of $P_S$ implies that
other quantities should have also an universal power law behavior with predictable exponents. This hypothesis is
tested using a numerical model of the phase space density of cold dark matter haloes, an excellent agreement is obtained. 
\end{abstract}
\maketitle
\section{Introduction.}
The formation of dark matter haloes in cosmology has been investigated in great details and impressive accuracy by numerical
simulations.
An important discovery was the existence of the universal NFW density profile (Navarro, Frenk, \& White, 1997). 
Later
Taylor \& Navarro (2001) discovered another intriguing property, the so-called pseudo-phase density, $\frac{\rho}{\sigma^3}$ where
$\rho$ is the projected density and $\sigma$ is the velocity dispersion, is a universal power law as a function of radius.
The exponent of this power law corresponds exactly to the exponent obtained for the self-similar solutions developped by Fillmore \& Goldreich (1984) in 
the special case of the secondary infall model studied by Bertschinger (1985).
{
The initial purely radial Bertschinger model was recently refined by Zukin \& Bertschinger (2010a, 2010b) by
introducing tidal torque. In this new approach Zukin \& Bertschinger confirmed the universality of the pseudo phase space density
at large and intermediate scale but questioned the universality at small scale. Very high resolution numerical simulations also raise questions about the
universality of the pseudo phase space density at small scale (Stadel {\it etal} 2009). An interesting addition to this debate
is the work of Nusser A. (2001) who show that the introduction of momenta does not modify the mass profile.
}
Recent investigations confirmed the universal power law behavior of the pseudo phase space density,
and improved the numerical accuracy by taking into account the ellipticity of the haloes and the substructures.
{
As an example Ludlow {\it etal} (2010) demonstrated that
the residual to the fit of a power law to the pseudo phase space density is typically about 10 to 20\% and does not exceed 30\%.
}
The power law behavior of the pseudo phase space density is now observed with impressive accuracy over 3 decades in radius. 
{
Note that the universality of power laws for the pseudo-phase density is stronger than the universality of the NFW profile (see Vogelsberger, Mohayaee, \& White 2011).
For a general review of the universal and not so universal properties of dark matter haloes, see Navarro {\it etal} (2010).
}
The relation
with the Bertschinger solution and the strong power law behavior of the pseudo phase space density seems to indicate a relation to
the self similar solutions of the Vlasov Poisson system. This paper will explore the details of the relation between the collapse of
a dynamically cold system in a cosmological context and self similarity. Before tackling the general problem of the multi dimensional
solutions, some attention will be given to a simpler problem, the one dimensional cold collapse. In particular it is interesting to consider
and study another universal behavior, the convergence of the one dimensional cold collapse to a universal power
law density with exponent $-\frac{1}{2}$ for a large class of initial conditions (Binney 2004).

\section{The Vlasov-Poisson system.}
The Vlasov-Poisson system for a space space density $f(\bf x, \bf v, t)$, with general space coordinates ${\bf x}$  velocity ${\bf v}$, time t, potential
$\phi(\bf x)$ and density $\rho(\bf x)$, is:
\begin{eqnarray}
\label{eq_self_1.1}
& \frac{\partial f}{\partial t}+\frac{\partial f}{\partial {\bf x}} {\bf v} -\frac{\partial f}{\partial {\bf v}} \frac{\partial \phi}{\partial {\bf x}} = 0\\
\label{eq_self_1.2}
&\Delta \phi= q_n G \rho \\
\label{eq_self_1.3}
& \rho = \int f \ d^n v
\end{eqnarray}
with $q_n$ a constant.
\section{Scale free solutions.}
The scale free solutions should be consistent with the following condition:
\begin{equation}
 f(\lambda_1 {\bf x},\lambda_2 {\bf v},\lambda_3 t )=\lambda_4 f({\bf x},{\bf v}, t )
\end{equation}
Considering the special case, $\lambda_i=1+\epsilon {\rm d} \lambda_i$, and expanding to
the first order in $\epsilon$, the following equation is obtained:
\begin{equation}
 \frac{\partial f}{\partial {\bf x}} {\rm d} \lambda_1 {\bf x} + \frac{\partial f}{\partial {\bf v}} {\rm d} \lambda_2 {\bf v}
+ \frac{\partial f}{\partial t} {\rm d} \lambda_3 t-f  {\rm d} \lambda_4=0
\label{eq_self_3}
\end{equation}
The solution to Eq \ref{eq_self_3} reads:
\begin{equation}
 f({\bf x}, {\bf v}, t) =f_0 t^{\alpha_0} F\left(\frac{{\bf x}}{t^{\alpha_1}},\frac{{\bf v}}{t^{\alpha_2}} \right)
\label{eq_self_4}
\end{equation}
with:
$$
 \alpha_0=\frac{{\rm d} \lambda_4}{{\rm d} \lambda_3} , \ \
 \alpha_1=\frac{{\rm d} \lambda_1}{{\rm d} \lambda_3} , \ \
 \alpha_2=\frac{{\rm d} \lambda_2}{{\rm d} \lambda_3}
$$
%
%
\section{Scale free solution of the Vlasov equation.}
The self similar solutions of the Vlasov equation are found by inserting  Eq \ref{eq_self_4} in Eq \ref{eq_self_1.1}.
We define the following scaled variables:  ${\bf x_2}=\frac{{\bf x}}{t^{\alpha_1}}$,
${\bf v_2}=\frac{{\bf v}}{t^{\alpha_2}}$. 
Using the former notations, a solution of the type given in
 Eq \ref{eq_self_4} is a solution of the Vlasov equation if:
\begin{eqnarray}\nonumber
&& (2+n \alpha_2) F+(1+\alpha_2)\frac{\partial F}{\partial {\bf x_2}} {\bf x_2} +\alpha_2 \frac{\partial F}{\partial {\bf v_2}} {\bf v_2} \\
&& -\left( \frac{\partial F}{\partial {\bf x_2}} {\bf v_2} 
 +\frac{\partial \tilde \phi}{\partial {\bf x_2}}\frac{\partial F}{\partial {\bf v_2}} \right)=0
\label{Self_Eq0}
\end{eqnarray}
and:
\begin{equation}
 \alpha_0=-2-n\alpha_2  , \ \ \alpha_1=1+\alpha_2
\label{Self_Eq}
\end{equation}
The definition of the scaled potential reads: 
\begin{equation}
\tilde \phi({\bf x_2})={t}^{-2 \alpha_2} \phi({\bf x})
\label{Scaled_pot}
\end{equation}
Note that an equation similar to Eq \ref{Self_Eq} was already derived by Lancellotti and Kiessling (2001).
\section{One dimensional solutions.}
The one dimensional solution ($n=1$ in Eq \ref{Self_Eq0}) corresponds to a two dimensional phase space described by coordinates ($x$,$v$), and
and associated self similar coordinates ($x_2$,$v_2$).
The solutions of Eq \ref{Self_Eq0} will be investigated for a power law potential, 
{
other quantities like the force and the density are also assumed to be power laws.
}
\begin{equation}
 \tilde \phi(x_2)= k |x_2|^{\beta+2}
\label{potential}
\end{equation}
The variable $x_2$ is re-scaled so that $k=\frac{1}{2}$.
.The  following change of variable $u=|x_2|^\eta$ is introduced with $\eta={\frac{\beta}{2}+1}$ in Eq \ref{Self_Eq0}. It is assumed that $-2<\beta<0$, as a consequence
the exponents of $\tilde \phi$ and $u$ are positive. 
{
But note that this requirement is not sufficient since the divergence of the total mass must be avoided. For $\beta<-1$ a finite mass
would imply a cutoff of the power law density at the center, and would generate a force which would not be a power law anymore.
This additional requirement implies that $-1<\beta<0$ for this type of one dimensional solutions.
}
The corresponding equation for the variable $u$ is:
\begin{eqnarray}\nonumber
 && 2+\alpha_2+(1+\alpha_2) \eta  \frac{\partial G}{\partial u} u +\alpha_2 \frac{\partial G}{\partial v_2} v_2 \\
            &&        -\sgn(x_2) \eta \ \left(\frac{\partial G}{\partial u} v_2 -2 k u \frac{\partial G}{\partial v_2}\right) u^{\frac{\beta}{\beta+2}}=0
\label{Eq_self_u}
\end{eqnarray}
where $\sgn(x)$ is the function:
$$
\sgn(x) = \left\{
    \begin{array}{ll}
        -1 & x < 0 \\
        1 & x \ge 0 \\
    \end{array}
\right.
$$
and $\log \left( F(x_2,v_2) \right)=G(u,v_2)$
We introduce again new variables, $(R,\psi)$, with:
$$
    \begin{array}{ll}
        R&=\sqrt{u^2+v_2^2} \\
        \cos \psi &=\left\{ 
                   \begin{array}{ll}
                        & \frac{u}{R} \ \ \ x_2 \ge 0 \\
                        & -\frac{u}{R}  \ \ x_2 < 0 \\ 
                   \end{array}
             \right.
    \end{array}
$$
The new equation in these variables is:
\begin{eqnarray}\nonumber
&& - \eta_2 \cos \psi \sin \psi (1+\alpha_2) \frac{\partial H}{\partial \psi}+R\left( \cos^2 \psi (1+\alpha_2) \eta_2+\alpha_2\right) \frac{\partial H}{\partial R} \\
&& +\eta \ \frac{\partial H}{\partial \psi} \left( R |\cos \psi| \right)^{\frac{\beta}{\beta+2}}+2+\alpha_2=0
\label{Eq_self_rpsi}
\end{eqnarray}
where $\eta_2=\eta-\frac{\alpha_2}{1+\alpha_2}$, and $H(R,\psi)=G(u,v)$ \\\\
An interesting case is when the potential $\phi(x)$ is stationary, which corresponds to, $\beta \alpha_1=-2$ and to $\eta_2=0$ 
{
(see Eqs \ref{Self_Eq} and \ref{Scaled_pot}).
} 
In this case Eq \ref{Eq_self_rpsi} is reduced to:
 \begin{equation}
2+\alpha_2 + R \alpha_2 \frac{\partial H}{\partial R} +
\frac{\alpha_2}{1+\alpha_2} \  \frac{\partial H}{\partial \psi} \left( R |cos(\psi)| \right)^{-\frac{1}{\alpha_2}}=0
\label{Eq_self_rpsi_eq}
\end{equation}
 Eq \ref{Eq_self_rpsi_eq} has a general solution:
\begin{equation}
H(R,\psi)=-\frac{\alpha_2+2}{\alpha_2} \log R+Q\left (R^{-\frac{1}{\alpha_2}} +\frac{1+\alpha_2}{\alpha_2 } \int |\cos(\psi)|^{\frac{1}{\alpha_2}} d\psi \right)
\label{sol_gen}
\end{equation}
Eq \ref{Eq_self_rpsi} has an asymptotic solution for small $R$ (and all values of $\eta_2$), the term in $R^{\frac{\beta}{\beta+2}}$ that multiplies the $\psi$ derivative of $H$ 
dominates other derivatives of $H$. The solution is:
\begin{equation}
 H(R,\psi)=-\frac{(\alpha_2+2)}{\frac{\beta}{2}+1} R^{-\frac{\beta}{\beta+2}} \int |\cos \psi|^{-\frac{\beta}{\beta+2}} d\psi + H_2(R)
\label{sol_R0}
\end{equation}
For a stationary potential, which corresponds to $\beta \alpha_1=-2$, Eq \ref{sol_R0} is reduced to:
\begin{equation}
 H(R,\psi)=-\frac{(\alpha_2+1)(\alpha_2+2)}{\alpha_2} R^{\frac{1}{\alpha_2}} \int |\cos \psi|^{\frac{1}{\alpha_2}} d\psi + H_2(R)
\label{sol_R01}
\end{equation}
Note that developing the general solution in Eq \ref{sol_gen} for $\eta_2=0$, small $R$, and adopting $Q(s)=-(2+\alpha_2) \ln(s) + Q_2(s)$ we obtain:
$$
\begin{array}{ll}
 & H(R,\psi) \simeq  Q_2(R^{-\frac{1}{\alpha_2}}) +\\
 & \indent \frac{1+\alpha_2}{\alpha_2} \int |\cos \psi|^{\frac{1}{\alpha_2}} d\psi \left( -{(\alpha_2 +2)} R^{\frac{1}{\alpha_2}} +Q_2^{'} (R^{-\frac{1}{\alpha_2}}) \right)
\end{array}
$$
Provided that $Q_2^{'} (R^{-\frac{1}{\alpha_2}})$ is of lower order than $R^{\frac{1}{\alpha_2}}$ the former expression reduces to Eq \ref{sol_R01}. Note that if $Q_2$ behaves
at origin like a power law with negative exponent this condition will be satisfied, and that as a consequence the behavior of the $Q_2$ solution at origin will not be singular, which
is a condition for the validity of the approximation at small $R$. As a consequence the general solution and the approximation at small $R$ are
consistent for $\eta_2=0$.   \\\\
%
%
%
 \section{Geometry of the solution.}
To study the geometry of the solution the iso-contours of the solution will be investigated. Considering a contour with phase
space density $F_0$, and using Eq \ref{sol_gen}, the equation of the corresponding contour is:
\begin{equation}
  \bar Q \left(R^{-\frac{1}{\alpha_2}}+I(\psi)  \right)=R^{\frac{\alpha_2+2}{\alpha_2}} F_0
 \label{fold_eq}
 \end{equation}
With the following definitions:
\begin{equation}
 Q=\ln \bar Q  \ \ {\rm and} \ \ I(\psi)=(1+\alpha_2)  \int |\cos(\psi)|^{\frac{1}{\alpha_2}} d\psi,
 \label{def_fold}
\end{equation}
 it is easy to realize that Eq \ref{fold_eq} corresponds to the equation of a spiral in phase space. Since
 $I(\psi) \simeq k \psi$ where $k$ is a constant, the equation reduces to $R=s(\psi)$, a specific
 example is a logarithmic spiral which would occur if $s$ is a power law plus a constant. An important
 parameter of this spiral structure is the inter-fold distance. This parameter is straightforward to measure in numerical
 simulations, and can be easily compared to the analytic expectation. The variation of $\psi$ between two consecutive folds is $2\pi$,
 thus if the position on one fold is $\psi_0$ the corresponding position on the consecutive fold is $\psi_1=\psi_0+2 \pi$.
 We will assume that the folds corresponds to a large number of turns, consequently $\psi_0 \gg 2 \pi$,
 and $I(\psi_1) = I(\psi_0) + \int_{0}^{2 \pi} |\cos(\psi)|^{\frac{1}{\alpha_2}} d\psi=I_0+\epsilon I_1$, with
 $\epsilon \ll 1$. Similarly we write the inter fold variation of $R_1=R_0+\epsilon dR$. Inserting these expressions
 in Eq \ref{fold_eq}, and using the original equation we obtain:
\begin{equation}
 \frac{dR}{R_0}= \frac{P(R_0^{\frac{\alpha_2+2}{\alpha_2}} F_0) I_1 \alpha_2}{P(R_0^{\frac{\alpha_2+2}{\alpha_2}} F_0) R_0^{-\frac{1}{\alpha_2}}+\alpha_2+2}
\label{dr_eq}
 \end{equation}
With:
\begin{equation}
 P(u)=\frac{\bar Q^{'} \left(\bar Q^{-1}(u) \right)}{u}
\label{P_eq}
\end{equation}
Requiring that all functions are power law's and starting with $Q$, then Eq \ref{P_eq} implies that $P$ is also a power law.
If $\frac{dR}{R}$ is also a power law Eq \ref{dr_eq} implies $P(u) \propto u^{\frac{1}{\alpha_2}}$, and as a consequence
we have:
\begin{equation}
 \frac{dR}{R_0}=\frac{F_0^{\frac{1}{\alpha_2+1}}  I_1 \alpha_2}{F_0^{\frac{1}{\alpha_2+2}}+\alpha_2+2} R_0^{\frac{1}{\alpha_2}} \propto R_0^{\frac{1}{\alpha_2}}
\label{fold_pos}
\end{equation}
An interesting consequence of Eq \ref{fold_pos} is the prediction of the projected density caustics. Introducing $R=x^{\frac{\beta}{2}+1}$ and
$\alpha_2=-\frac{2}{\beta}-1$, the position ${\rm p_n}$ of n$^{\rm th}$ fold is:
\begin{equation}
 p_n=n^{\frac{2}{\beta}}
\label{caust_pos}
\end{equation}
The self similar solution of Bertschinger (1985) corresponds to $\beta=-\frac{9}{4}$ which implies $p_n=n^{-\frac{8}{9}}$.
Sikivie \& Kinney (2001) proposed the following numerical approximation, $p_n \simeq \frac{1}{n}$ which is quite close
to the prediction of Eq \ref{caust_pos}. 
{
For more details about the formation of caustics and the fine grain structure 
of phase space in the cold dark matter model see, Mohayaee {\it etal} (2006),
Afshordi, Mohayaee \& Bertschinger (2008), Duffy \& Sikivie (2008), 
Vogelsberger {\it etal} (2008), White \& Vogelsberger (2009), Vogelsberger \& White (2011).
}
It is important to note that the thickness of a given contour can be derived using the same method as
the inter-folds distance. The variation of position for a contour in Eq \ref{fold_eq} corresponds to a variation
of $I(\psi_1) = I(\psi_0+d\psi)$, where $d\psi$ corresponds to a differential in the initial conditions for two opposite
sides of the contour. As a consequence the calculations are identical to the calculations performed for the inter-fold
distance, and the contour thickness is proportional to the inter-fold distance.
\section{Multi dimensional solutions.}
It is useful to define the self similar coordinates in the 6D space, $({\bf x_2},{\bf v_2})$, and the corresponding
modulus of distances $r$ and velocities $v$. The solution in several dimensions is constructed using the results obtained in one dimension.
{
Note that in three dimension $\beta>-1$ is not required anymore, to avoid the divergence of the mass at center $\beta>-3$ is sufficient.
}
The solution is the sum of a general solution plus a special solution. To facilitate the calculations we introduce
 $G({\bf x_2},{\bf v_2}) = \log F({\bf x_2},{\bf v_2})$, and a power law potential $\tilde \phi(r)=\frac{1}{2} r^{\beta+2}$, $-2<\beta<0$,
 Eq \ref{Self_Eq0} then reads:
\begin{equation}
 \begin{array}{ll}
& (2+n \alpha_2) +P_1+P_2=0 \\
& P_1=(1+\alpha_2)\frac{\partial G}{\partial {\bf x_2}} . {\bf x_2} +\alpha_2 
\frac{\partial G}{\partial {\bf v_2}} . {\bf v_2}  \\
& P_2=- \frac{\partial G}{\partial {\bf x_2}} . {\bf v_2}
 +(\beta+2) r^{\beta+1}\frac{\partial G}{\partial {\bf v_2}} . {{\bf x_2}}
 \end{array}
\label{Multi_dim_eq}
\end{equation}
considering a solution of the type:
\begin{equation}
G({\bf x_2},{\bf v_2})=K_0(R)+K_1(r,v_r)
\label{sol_multi}
\end{equation}
Here $R=\sqrt{v^2+r^{\beta+2}}$, and $K_1(r,v_r)$ is a function of the distance modulus and radial velocity $v_r$. The special
solution $K_1$ is constructed using the method we developed for the one dimensional solution. Provided we substitute the $(|x|,v)$
space with the $(r,v_r)$ space the equations are the same, and finding the solution $K_1$ is thus a problem we have already solved. Assuming
that $K_1$ is a full solution or an asymptotic solution for small $R$, we introduce the solution Eq \ref{sol_multi} in Eq \ref{Multi_dim_eq}.
Thus considering  $G=K_0+K_1$ the only remaining significant terms may come from $K_0$ only. Since basically $K_0$ is a function of energy, we have $P_2=0$.
Using the $R$ variable, we find that $P_1$ behaves like $R \frac{\partial K_0}{\partial R}$. Since the terms relative to $K_1$ are of order of unity,
the $P_1$ term of $K_0$ must be compared to unity. Provided that the function $K_0$ behaves like a power law with positive exponent,  
the relevant $P_1$ term will become negligible at small $R$ and the solution in Eq \ref{sol_multi} is an asymptotic solution at small $R$.
\section{Pseudo phase space density for the three dimensional solution.}
{
The power-law behavior of the pseudo phase space density observed in 3D numerical simulations is a strong motivation to study
 the pseudo phase space density of the solutions.
  For a comparison to the numerical simulations the quantity of interest is the
 smooth pseudo phase space density  (see Sec. 9 for a definition of the smooth quantities). 
  It is simple to construct an analytic self similar solution at equilibrium by using a power law model. 
 Provided all histograms of the values
 of $f$ are power laws, the smoothed density of $f$ will also correspond to a power law. For an equilibrium model it is straightforward to derive
 the parameters of this power law model. However due to the geometric nature of the solution (see Sec. 6) 
  the cold solution occupies a finite range. The cold solution is a series of folds,
 with the first fold
 as the boundary of the system. Sec. 6 shows that this outer boundary may be approximated with a constant radius in $R$, $R=R_0$. As a consequence the density
of the cold solution and of its smoothed expectation must be cut outside some radius $R_0$. The smooth pseudo phase density for the cold
density is thus obtained by estimating the expectation for the smooth power law model with initial density $r^{-\beta_0}$ within $R_0$. The variable $r$
is the modulus of ${\bf x}$, at this point it is also useful to introduce the variable $v$, the modulus of ${\bf v}$.
In three dimension the pseudo phase density is $\frac{\rho}{\sigma^3}$,
thus the following integrals have to be estimated numerically:
$$
\rho=\int_{0}^{v_0} (\frac{v^2}{2}+\phi)^{\eta} v^2 dv
$$
$$
\sigma^2=\frac{\left( \int_{0}^{v_0} (\frac{v^2}{2}+\phi)^{\eta} v^4 dv \right)}{\rho}
$$
With:
$$
v_0=\sqrt{2 (R_0^2-\phi)}
$$
$$
\eta=-\frac{1}{2} \frac{\beta_0+6}{\beta_0+2}
$$
And:
$$
\phi= \phi_0 r^{\beta_0+2}
$$
note that in the above integrals the re-scaling of $R_0$ implies the re-scaling of $r$ and the re-scaling of the integrals, thus the general shape
 of $\frac{\rho}{\sigma^3}$ may be obtained for any value of $R_0$. The results of the calculations are presented in Fig \ref{pseudo1}.
 \begin{figure}
\includegraphics[angle=0,scale=.435]{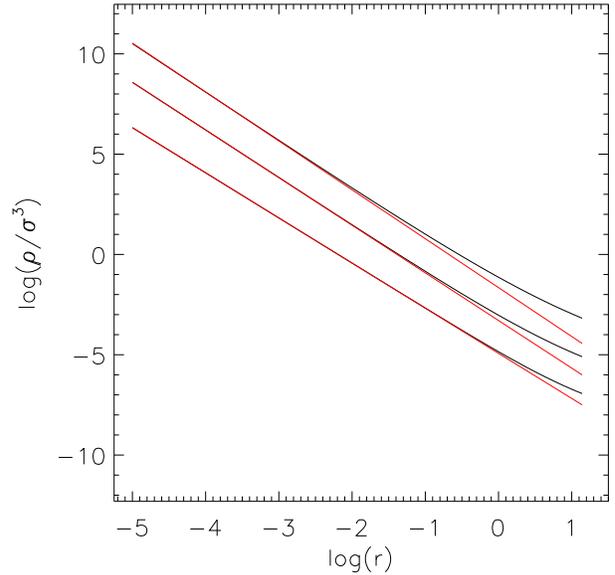}
\caption{The general shape of the smoothed pseudo phase space density for cold 3D self-similar solutions. The pseudo phase space density was obtained
by integrating the power law expectation within a boundary, see Sec. 8 for more explanation. The variations of $\frac{\rho}{\sigma^3}$ are represented 
 from top to bottom for $\beta_0=-\frac{9}{8}$, $\beta_0=-\frac{5}{4}$, $\beta_0=-\frac{3}{2}$.}
\label{pseudo1}
\end{figure}
The pseudo phase density of the solution is consistent with the theoretical power law expectation near the center and deviates near the edge of the system.
This is obviously expected since the cut has more effect at the edge than at the center. An interesting consequence of the behavior of these
solutions, is that even if the pseudo phase density deviates significantly from a power law near the edge of the system it is still an auto-similar
solution in this area. Although the solution computed
here is isotropic which is not the case of the dark matter haloes investigated in numerical simulations, this result suggests that even if the pseudo phase space density
is not a power law in some area, it is still possible that the phase space density is self-similar in this area.
}
\section{One dimensional cold collapse.}
\subsection{Power law regime in the central region.}
\begin{figure}
\includegraphics[angle=0,scale=.435]{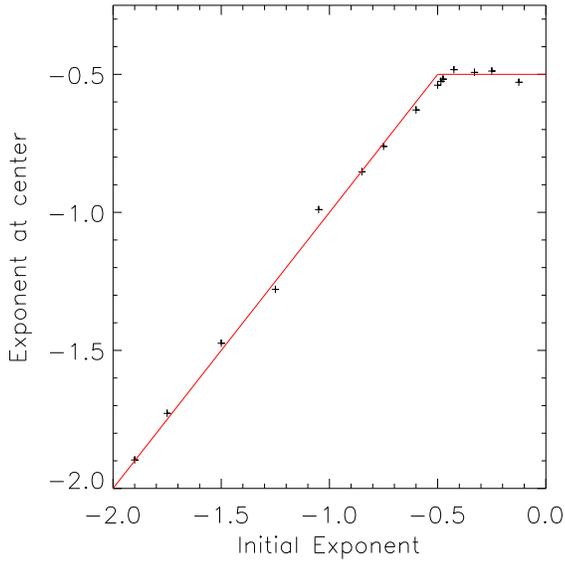}
\caption{
A power law is fitted 
to the initial density and to the final late time density obtained after a large number of orbits.
The resulting exponents are plotted as a function of each other for various initial conditions. 
For an initial exponent $\beta_0<-\frac{1}{2}$ the final exponent is identical, but for $\beta_0>-\frac{1}{2}$ 
a break occur and the final exponent saturates to $-\frac{1}{2}$, see text for more explanations.
Note that in the range $0<\beta_0<-1$ the exponent was evaluated by fitting a power law to the force, and making
the necessary conversion.
}
\label{pow0}
\end{figure}

\begin{figure}
\includegraphics[angle=0,scale=.435]{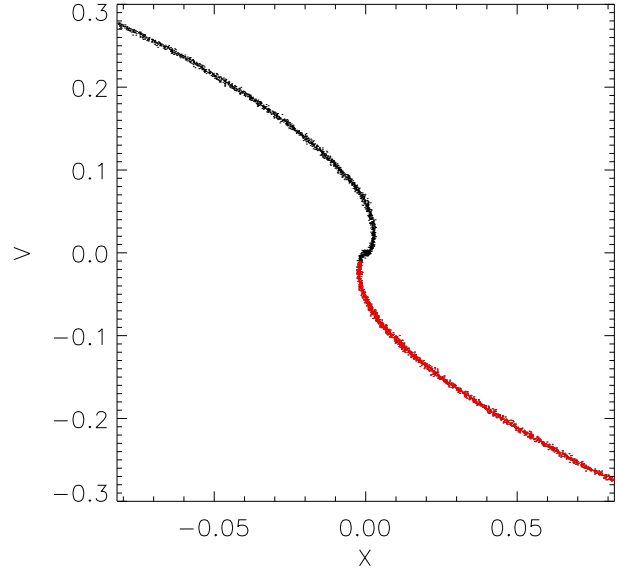}
\caption{The phase space structure in the central region of the system just after the first crossings of the particles. 
This simulation contains a total of $4 \times 10^5$  particles. The initial conditions are dynamically cold 
with projected density profile $x^{-\frac{1}{3}}$ and Gaussian velocity dispersion.}
\label{pow2}
\end{figure}

\begin{figure}
\includegraphics[angle=0,scale=.435]{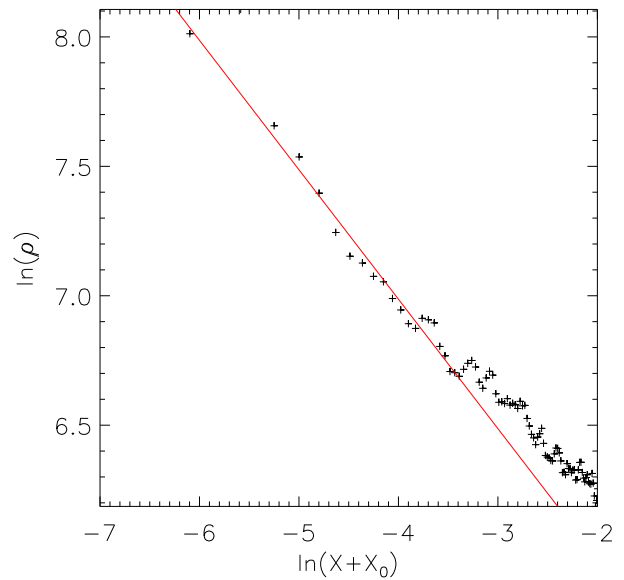}
\caption{The projected density associated with the first caustic. The first caustic corresponds to the red points in Fig \ref{pow2}. To demonstrate the relation
between the power law projected density and the projected density associated with the caustic, the $x$ coordinate system was re-centered on the caustic 
position by introducing a shift in position $x_0$. The red-line represents a power law with exponent $-\frac{1}{2}$.}
\label{pow1}
\end{figure}

 The evolution of a system of particles with small initial velocity dispersion in the two dimensional
 phase space under the action of one dimensional planar gravity will be investigated in this section. Interestingly Binney, (2004) found that
 for such systems the density near the center is very close to $x^{-\frac{1}{2}}$. This power law density
 suggests the possibility of a universal self-similar regime in the central region.  Binney (2004) proposed that the  $x^{-\frac{1}{2}}$
 regime is valid for a very large class of initial conditions, but an interesting problem is to find for what type
 of initial conditions the $x^{-\frac{1}{2}}$ is not obtained, and where is the limit. This problem may be solved by using the following
 simple approach, we take a set of cold initial conditions with power law densities and compute the density
 of the system at late times. 
{
The initial conditions are sampled by using $4 \times 10^5$ particles, the phase space coordinates ($x$,$v$) of these particles are initiated using
random number generators. The $x$ positions are initiated using
a power law random generator, while a Gaussian random generator is used for the $v$ coordinates.
The $x$ dimension random number generator is constructed by applying a power law transform to the output of a uniform random generator.
The initial conditions are evolved using a leapfrog integrator (Birdsall \& Langdon, 1985). The leapfrog is a symplectic integrator which is particularly
important for reconstructing the structure of phase space. The leapfrog method requires the evaluation of the force at the position of each particle which is computed by subtracting
the total mass on each side of the particle. No smoothing is applied to the force. The final step is to evaluate the proper time step for the leapfrog integration.
The time step must be small enough to allow a proper reconstruction of the shortest orbits in the system. The shortest orbits are the closest to the center of the system.
Assuming that the typical size of the system is unity if we require a dynamical range in $x$ of $D_0$, the associated shortest distance is $x_0=\frac{1}{D_0}$. In the pre-collapse
regime the fall time of a particle $T_0$ to the center depends on the initial mass $M_0$ inside the particle shell. In a system with initial power law density $\rho=\rho_0 x^{\beta_0}$,
$T_0=\sqrt{\frac{x_0}{G M_0}}=\sqrt{\frac{\beta_0+1}{G \rho_0}} x^{-\frac{\beta_0}{2}}$. The proper time step for leapfrog integration should be a fraction of $T_0$; 
numerical experiments
show that a fraction of $5\%$ is optimal. Another point is the estimation of the initial velocity dispersion $\sigma_0$. Basically the resolution
in the velocity space $\sigma_0$ has to match the resolution in $x$ space $x_0$. The velocity at the center of a particle with initial distance $x_0$ should
be of the order of $\sigma_0$, which corresponds to $\sigma_0 \simeq \sqrt{2 G M_0 x_0} = 2 \sqrt{\frac{G \rho_0}{\beta_0+1}} x_0^{\frac{\beta_0}{2}+1}$. Note that the 
above calculations of $T_0$ and $\sigma_0$ are not fully analytic for $\beta<-1$, in this case the mass $M_0$ was computed directly
from the initial conditions. 
}
%
%
 The results of the 1D numerical simulations are presented in Fig ~\ref{pow0}. { The resolution adopted corresponds to $D_0=1000$.}
{
 For a better numerical efficiency, in the range $-1<\beta_0<0$ the power law exponent
 of the force $\beta_F$ is computed, the force is much smoother than the density (affected by caustics), the equivalent projected density exponent, is $\beta_F-1$.
 For $-2<\beta_0<-1$ the power law exponent was computed directly by fitting a power law to the density.
}
 Interestingly the initial exponent $\beta_0$ is conserved at later time for $\beta_0<-\frac{1}{2}$, while for larger
 values of $\beta_0$ the final exponent has a constant value of $-\frac{1}{2}$ (see Fig \ref{pow0}). As a consequence the $x^{-\frac{1}{2}}$ density
 at the center is not universal but does occur precisely when $\beta_0=-\frac{1}{2}$. Note that this effect was also
 noticed by Schulz {\it etal} (2012) with results quite similar to Fig \ref{pow0}. This particular behavior suggests a cut-off, 
 no final exponent is larger than $-\frac{1}{2}$. To understand the cause of this cut-off it is interesting to investigate 
 numerical simulations with initial conditions $0>\beta_0>-\frac{1}{2}$ and look for the first appearance of the $x^{-\frac{1}{2}}$ power law.
 This analysis is illustrated for $\beta_0=-\frac{1}{3}$, the first appearance of the $x^{-\frac{1}{2}}$ projected density occurs 
 when the the first fold starts to form or equivalently just after the first crossing of particles with opposite velocities at
 the center. The phase space structure at this time is presented in Fig ~\ref{pow2}, we notice that near the center we have two
 caustics. These two caustics are very close to the center and according to the theory of singularities for a one dimensional density,
 we expect that their projected density will go like $\tilde x^{-\frac{1}{2}}$,
 where $\tilde x$ is a coordinate centered on the caustic. This is just what we obtain in Fig ~\ref{pow1}. The lower fold (red points in
 Fig ~\ref{pow2}) has a projected density close to  $\tilde x^{-\frac{1}{2}}$ near the center. This density starting from the caustics
 extends at distances significantly larger than the separation between the caustic and the center, which means that in a large portion
 of this regime $\tilde x \simeq x$ and that as a consequence the density goes like $x^{-\frac{1}{2}}$. The $\tilde x^{-\frac{1}{2}}$ regime
 assume that the phase space density along the fold is constant, but due to the high non-linearity at the crossing the stretching of the phase
 space is large which implies that the density along the fold is nearly constant in a significant domain. This density associated with the caustic
 initiate the $x^{-\frac{1}{2}}$ asymptotic regime at the center and since it is dominant for $\beta_0>-\frac{1}{2}$ it is prone to extend to larger values
 of $x$. At later times the system forms a large number of folds, with the persistence of the  initial caustic induced
 $x^{-\frac{1}{2}}$ regime at the center.
 The  $x^{-\frac{1}{2}}$ central regime is also
 present for other initial conditions, for instance, a Gaussian, 2 Gaussian,.., and many other, but the difference is that even if this regime is present very
 close to the center we find also a rather brutal transition to a tiny area with nearly constant density. This transition occurs at the very
 center of the system, and the size of this constant region depends on the initial conditions. This constant region at the center
 is probably a remain of the nearly flat density present in some initial conditions. The transition from the $x^{-\frac{1}{2}}$ regime
 to the constant density regime signal also a break of the self similar regime. In any case the existence 
 of this small constant region affect the extent of the $x^{-\frac{1}{2}}$ regime which is still present and most likely for the same reasons that
 for power law initial conditions.
 For a further investigation of this issue it is interesting to point the contribution of Colombi \& Touma (2012) on this particular
 topic.
 \subsection{Self similarity in the central region.}
\begin{figure}
\includegraphics[angle=0,scale=.435]{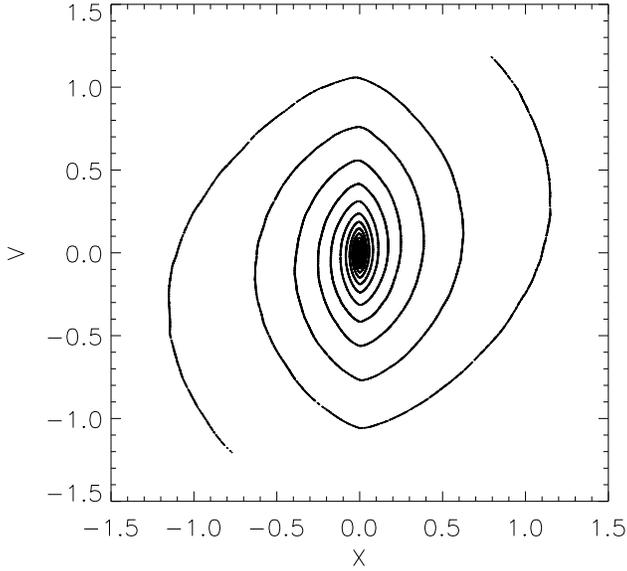}
\caption{The late time evolution of the phase space structure for the cold system.
The initial conditions correspond to the density profile 
$x^{-\frac{1}{3}}$  already presented in Fig \ref{pow2}. In this figure
the most visible folds are the first outer folds, see Fig \ref{pow4} for a more detailed view of the inner fold system.}
\label{pow3}
\end{figure}

\begin{figure}
\includegraphics[angle=0,scale=.435]{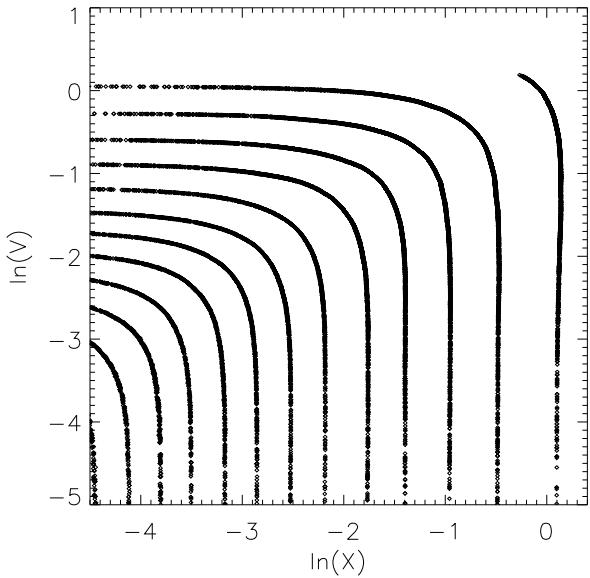}
\caption{This plot presents a detailed view of the upper quadrant ($x>0$, $v>0$) of the phase space diagram presented in Fig ~\ref{pow3}. Logarithmic coordinates
were introduced for both axis to allow a better view of the inner space folds.}
\label{pow4}
\end{figure}

\begin{figure}
\includegraphics[angle=0,scale=.435]{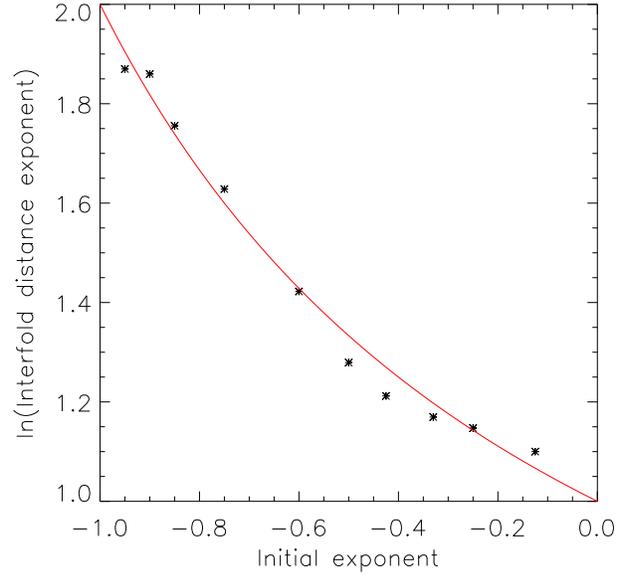}
\caption{The power law exponent $\gamma$ of the inter-fold distance $dR$. The exponent $\gamma$ was reconstructed by fitting the power 
law $dR=R^{\gamma}$ to the fold system in phase space. The variable $R$ is defined as: $R=\sqrt{v^2+\phi}$. 
The continuous line corresponds to the theoretical expectation (see Eq \ref{fold_pos}).}
\label{pow_f}
\end{figure}

\begin{figure}
\includegraphics[angle=0,scale=.435]{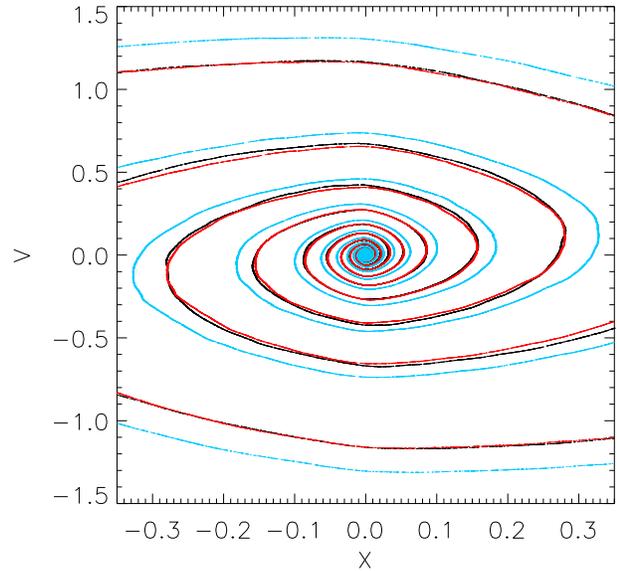}
\caption{The fold system at the center at two different times (black and blue).  The initial conditions corresponds to the density profile 
$x^{-\frac{1}{3}}$  already presented in Fig \ref{pow2}.
The blue folds are re-scaled using $\alpha_2=3$. The time-scale factor is equal to the ratio of the time from the beginning of the simulation
for each fold system.}
\label{self_plot}
\end{figure}

 It is important to note that in the central region of all
 numerical simulations self-similarity was observed (see Fig ~\ref{self_plot} for an illustration). 
 The exponent $-\frac{1}{2}$ of the density implies $\alpha_2=3$, which
 means that in the $x$ dimension the simulation should scale like $\frac{1}{t^3}$ and in the $v$ dimension like  $\frac{1}{t^4}$ . Using these
 scalings the folds at different times should be at the same location, which could be verified with good accuracy. This property definitely proves
 that the $x^{-\frac{1}{2}}$ corresponds to a self similar regime in phase space. Since analytic formula for the self-similar solution
 with power law density were derived in the former sections, it is interesting to check the consistency with the fold system.
 The fold system is illustrated in Figs ~\ref{pow3} and  ~\ref{pow4}.
 For power law initial conditions, it is expected that all quantities will be power
 laws, and thus the inter-fold distance should be given by Eq \ref{fold_pos}. The potential $\phi(x)$ was computed, as well as $R=\sqrt{v^2+\phi}$,
 and finally the inter-fold distance $dR$ was estimated. The results are presented in Fig \ref{pow_f}. The exponent of the inter-fold distance as a function
 of $R$ according to Eq \ref{fold_pos} is $\gamma=\frac{1}{\alpha_2}+1$ and provided that $\alpha_2$ is deduced from the initial condition exponent $\beta_0$,
 we have, $\alpha_2=-\frac{2}{\beta_0}-1$ and $\gamma=\frac{2}{\beta_0+2}$. Fig \ref{pow_f} shows that this theoretical expectation of $\gamma$ is in good agreement
 with the numerical estimations. 
%
%
 It is interesting to note that we expect that the theoretical expectation of $\gamma$ is going to match the data for $\beta_0 <-\frac{1}{2}$,
 but for  $\beta_0 >-\frac{1}{2}$ we would expect $\beta=-\frac{1}{2}$ and not $\beta=\beta_0$. This is not what is observed, and we must conclude that despite the
 $x^{-\frac{1}{2}}$ law at the center the structure of the fold system is preserved for $\beta_0<-\frac{1}{2}$. 
  Other type of initial conditions imply different inter-fold distance, but this
 is always a small correction from the $dR \propto R^\gamma$ law. 
 To conclude it is interesting to note that all solutions for a large class of initial conditions have 
 power law densities with exponents close to $-\frac{1}{2}$ at the center, and are all self-similar with the same similarity index at the center. This definitely indicates that
 the degree of freedom ($Q$ in Eq \ref{sol_gen} or $H_2$ in Eq \ref{sol_R0}) is important since it allows some flexibility to accommodate the initial conditions. All solutions
 have similar density at the center but have different phase space distributions while at the same time they remain self-similar.
\section{Properties of stationary self-similar solutions in multi dimensions.}
\begin{figure}
\includegraphics[angle=0,scale=.435]{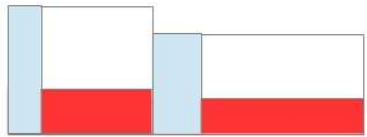}
\caption{The cold un-smoothed density is represented in blue and the local value of the smoothed density is represented
in red in this simple 1D analogy. The cold density is at the position of the fold and its thickness is proportional to the inter-fold distance.
The smooth density corresponds to the averaging of the cold density between two consecutive folds. Since the thickness and the inter-fold
distance are proportional, the construction of the smooth density is equivalent to a simple re-scaling of the cold density. Obviously 
the cold and smoothed density cannot be proportional in all points
of phase space, since the cold density is zero in some place, and the smooth density is neither equal to zero between folds. However
the distribution of the values of the smooth and cold density differs only by scale factors, which implies that the 
self similar (scaling) properties of the cold distribution of $f$ values are preserved in the smooth density distribution, or probability distribution of the smooth $f$.}
\label{smooth}
\end{figure}

\begin{figure}
\includegraphics[angle=0,scale=.435]{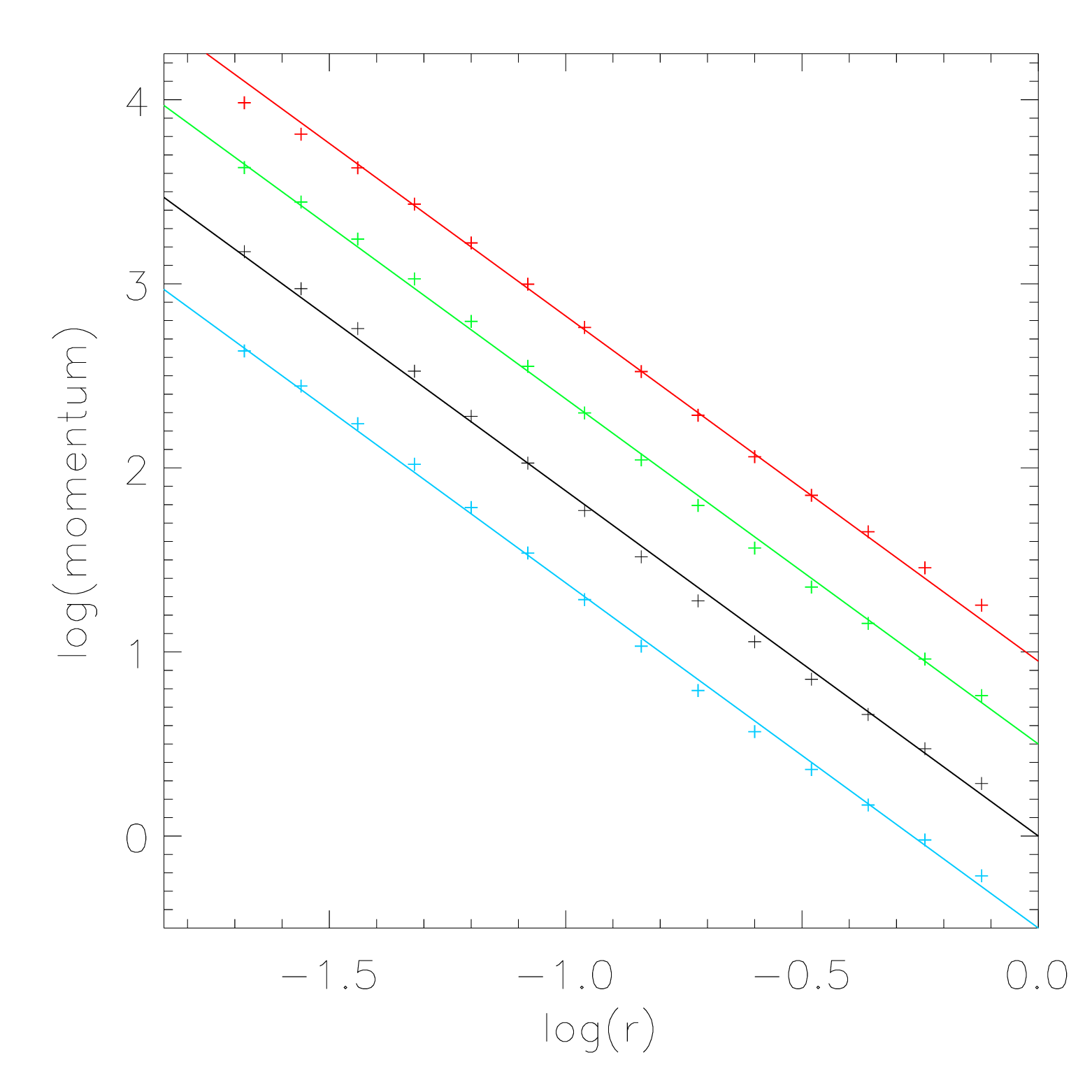}
\caption{The following quantities $\frac{\rho}{\sigma^3}$ black, $\left[\frac{\int f^2 d^3v}{\sigma^3} \right]^{\frac{1}{2}}$ blue, 
$\left[\frac{\int f^3 d^3v}{\sigma^3} \right]^{\frac{1}{3}}$ green,
$\left[\frac{\int f^4 d^3v}{\sigma^3} \right]^{\frac{1}{4}}$ red, are represented by points (cross). The points were calculated using
the Wojtak etal (2008) model. A straight line with slope $-\frac{15}{8}$ is over-plotted over each curve.}
\label{mam2}
\end{figure}

\begin{figure}
\includegraphics[angle=0,scale=.435]{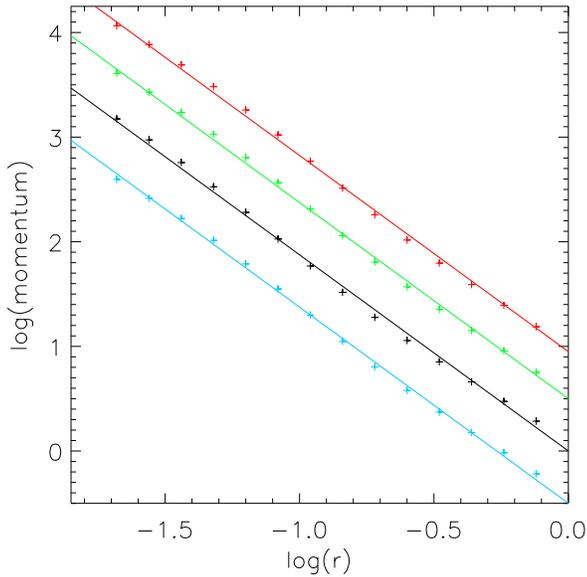}
\caption{The following quantities $\frac{\rho}{\sigma^3}$ black, $\frac{\int f^2 d^3v}{\int f d^3v}$ blue, 
$\left[\frac{\int f^3 d^3v}{\int f d^3v} \right]^{\frac{1}{2}}$ green,
$\left[\frac{\int f^4 d^3v}{\int f d^3v} \right]^{\frac{1}{3}}$ red, are represented by points (crosses). 
 The points were calculated using the Wojtak etal (2008) model.
A straight line with slope $-\frac{15}{8}$ is over-plotted over each curve.}
\label{mam1}
\end{figure}

The collapse of a dark matter halo in cosmological conditions is a very different problem from the 1D collapse investigated in the former section.
{
The relation between the three dimensional collapse and self similarity was already explored by Lithwick \& Dalal (2011).
}
Here self similarity is imposed by the cosmological infall, the self similarity is not a consequence of a singularity at the center or of power law
initial conditions. 
In cosmology the initial stage of the halo formation can be represented with the accretion of matter
on a seed mass. This process implies that the turn-over radius of the in-falling shells of matter
is a power law of time (Gunn 1977, Bertschinger, 1985). The relevant similarity exponent in the notations of this paper
is $\alpha_2=-\frac{1}{9}$. The self similarity of the solution is enforced by the cosmological
infall, which implies that the final solution should reflect the similarity of the infall (Taylor \& Navarro 2001). Thus
we have good reasons to assume that the final outcome of the collapse of a dark matter halo will
correspond to a self-similar solution of the Vlasov-Poisson system. Since dark matter is cold
the solution will correspond to the multiple folds of the initial surface in phase space.
However, an equilibrium or a near equilibrium situation is a very different problem for a smooth continuous
system in phase-space and for the folding of a cold system in phase space. Cold systems forms spiral
in phase space and even in their late evolution they will be still evolving and will not converge to a smooth continuous
 equilibrium system. The smoothing of the spiral will produce a smooth phase-space density, and that smoothed density
will be nearly stationary at late times. However in general the problem is that
the smoothed density is not necessarily auto-similar. The smoothed density will correspond to the smooth
auto-similar solution only in some special cases. This problem is due to the fact that the smoothed phase
space density does not satisfy the ordinary Vlasov equation but a different differential equation. Since
we cannot relate the cold self-similar solutions to the smooth self-similar solutions we cannot benefit
from the fact that the smooth equilibrium solutions are simple analytic functions (Lancellotti \& Kiessling, 2000). 
However it is clear that at late time the cold self similar solution should have a nearly stationary smoothed
density in phase space and that it will correspond to a given sub-set of self similar solutions. The final
problem is thus to identify the properties of this sub-set of solutions. A key point in the solution of this issue
is to observe that even if the cold solution is discontinuous in phase space, the probability distribution
of the phase space density $P(f)$ could be similar to that of a smooth solution $P_S(f)$. 
Actually if we consider the local surfaces of folds in phase space within some small solid angle, 
the integrated density between two local surfaces would correspond to the
local surface multiplied by the thickness of the fold. The integration and averaging of the density between two folds corresponds
to the smoothing operation. If the fold thickness is proportional to the distance between the folds, then the smoothed
density will be proportional to the cold un-smoothed density, and its occurrence will also be proportional to the cold occurrence since it is
 proportional to the ratio between the thickness and the inter-fold distance. This reasoning shows that the smoothed probability density
 of $f$ can be deduced from the un-smoothed density by the introduction of constant scale factors (see Fig \ref{smooth} for an illustration). 
 As a consequence the self-similar properties
 of the cold probability distribution will be also valid for the smooth probability distribution. Note that $f$ is made of many different contours, but since
 the scaling properties between $P$ and $P_S$ are valid for each contour, the transfer of self similarity between $P$ and $P_S$ remain valid
 in presence of many different contours.
 The property 
 of proportionality between the thickness and the inter fold distance was demonstrated in Sec. 6, but
 is also required for the existence of the solution.
If the thickness exceeds the inter fold distance then a crossing would occur near the origin, while if the opposite behavior happened
the crossing would happen on larger distances. The only solution is thus that the fold thickness is proportional to the fold inter distance.
As a consequence it is clear that the smooth probability distribution inherit the self similar properties of the cold solution, and that
power-law integrated quantities estimated from $P_S(f)$ should have an auto-similar behavior. Note that a priori the equivalence
between $P(f)$ and $P_S(f)$ is valid on any sub domain of the phase space, provided that this domain is large enough
to define the statistics of $f$. In particular all quantities like:
$$
Q_n({\bf x_0})=\int P_S(f({\bf x,v})) f({\bf x,v})^n df 
$$
$$
{\rm with:} \ \ {\bf x=x_0} \ \  {\rm and:} \ \ 0 <v< \infty
$$
have to be self similar since 
$$
Q_{n}^{0}({\bf x_0})=\int P(f({\bf x,v})) f({\bf x,v})^n df 
$$ 
$$
{\rm with:} \ \ {\bf x=x_0} \ \  {\rm and:} \ \ 0 <v< \infty
$$
Is self-similar and $P_S$ inherits the self similarity of $P$. As a consequence the smooth self-similar $Q_n$ function
should write as a power law of time multiplied with a function of ${\bf x_2}$. 
A direct estimate of the mean value of $f$ $Q_1$ is:
$$
 Q_1({\bf x})=\int P_S(f) f df \simeq \frac{1}{8 n^3 \sigma^3}\int_{-n \sigma}^{n \sigma} f d^3v  \propto \frac{\rho}{\sigma^3}
$$
In the same spirit one may approximate $Q_n$:
$$
 Q_n({\bf x}) \propto \frac{\int f^n d^3v}{\sigma^3}
$$
The ratio of $Q_n$ quantities is also self-similar:
$$
R_{nm}=\frac{Q_n}{Q_m}
$$
Since $Q_n$ and $R_{nm}$ are smooth continuous functions and self similar:
$$
 Q_n({\bf x})=t^{n \alpha_0} h_1({\bf x_2})
$$
and
$$
 R_{nm}({\bf x})=t^{(n-m) \alpha_0} h_2({\bf x_2})
$$
Stationary solutions for $Q_n$ and $R_{nm}$ implies that $h_1$ and $h_2$ are power law's, thus:
$$
 Q_n(r) \propto t^{n \alpha_0}  \left [ \frac{r}{t^{\alpha_1}} \right]^{\gamma_1}=r^{\gamma_1}  \ \ ; \ \ \gamma_1= n \frac{\alpha_0}{\alpha_1} = -n \frac{15}{8}
$$
Similarly:
$$
  R_{nm}(r) \propto t^{(n-m) \alpha_0}  \left [ \frac{r}{t^{\alpha_1}} \right]^{\gamma_2}=r^{\gamma_2}  \ \ ; \ \ \gamma_2=-(n-m)\frac{15}{8}
$$
(Note that we assume isotropic systems, thus ${\bf x}$ was replaced with the distance modulus $r$.) \\
The calculation of integrals involving higher order power of f ($f^n$, $n>1$) requires the knowledge of the spatial distribution of $f$.
The reconstruction of $f$ from numerical discrete numerical data is a difficult and uneasy task
 (See Arad, Dekel, \& Klypin, 2004). A more convenient way of estimating the higher order terms is to use
the  accurate analytic approximations of $f$ that were developed
 by Wojtak, {\it etal} (2008). All integrals were performed using Eq 29 from  Wojtak, {\it etal} (2008). The results
 are presented in Figs ~\ref{mam2} and ~\ref{mam1}. For convenience an inverse power law's of higher order quantities
 was taken to reduce all self similar exponents to the $-\frac{15}{8}$ theoretical self-similar expectation. 
 All quantities are nearly linear in the log-log
 diagram, and all slopes in this diagram are very close to $-\frac{15}{8}$. The only possible explanation is that
 if all moments of the smooth probability distribution $P_S(f)$ are self similar is that  $P_S(f)$ itself is self similar.
 As a consequence the universal behavior observed for CDM haloes in numerical simulations reflects the self similarity of $P_S(f)$. \\\\
\section{Conclusion and perspectives.}
The convergence of cold initial conditions with initial density profile not steeper that $x^{-\frac{1}{2}}$ to a universal
power law density with exponent $-\frac{1}{2}$ (the Binney conjecture, BC) is an interesting and powerful example of universality.
The one dimensional solutions are quite simple, and the BC is a very useful and efficient tool to explore
self similarity. The conjecture implies that a large class of self similar solutions forms at the center of collapsed
systems. The analytic solutions developed in this paper help to understand the  general properties of the solutions, and
the basic mechanism at the origin of the conjecture, which is related to a singularity has been identified.
However the full richness and complexity of the 1D self similar solutions have yet to be explored, and the BC provides
a very interesting method to generate a large class of self similar solutions. The final self similar solution
will depend on the initial conditions, and it is clear that more complicated and exotic initial conditions have to be explored.
The problem of the 3D cold collapse in a cosmological context, is somewhat different since the similarity is not due to a singularity
but is enforced by the nature of the infall. In 3D simulations of dark matter haloes the pseudo phase space density $\frac{\rho}{\sigma^3}$
is an universal power law. It was proved that this similarity is not the only one, and that other moments of the probability distribution
of $f$ have the same universal properties. The convergence to a power law of all these quantities is due to the self similarity of
the probability distribution of the smoothed $f$, which is inherited from the self similarity of the original distribution. Interestingly
it was shown by  Dehnen \& McLaughlin (2005) that the $\frac{\rho}{\sigma^3}$ universal power law implied universal density profiles, not exactly
NFW profiles, but quite close.
{
It is interesting to point that the NFW profile itself may be understood usimg simple physics
(Dalal, N., Lithwick, Y., Kuhlen, M., 2010).
}
As a consequence one may conclude that all the universality properties of the dark matter haloes are related
the self similarity of the probability distribution of the smoothed $f$, $P_S(f)$. A final point is that as expected, the self similar properties
of $P_S(f)$ are also present for the one dimensional solutions presented in this paper, and that higher order moment of $P_S(f)$ are also
power laws, which is a fundamental property of self similar solutions.
{
A fundamental and natural perspective is to extend this work to
real galaxies where not only dark matter is present but also baryonic matter. It is remarkable that the rotation curves of galaxies show also some universal
similarity properties (Donato {\it etal} 2009, Salucci {\it etal} 2007.)\\\\
}
\thanks{
The author would like to thank Scott Tremaine for support during his stay at the Institute for
advanced studies in 2011. The author thanks St\'ephane Colombi, Jean Philippe Beaulieu, and Jihad Touma for interesting comments.}
{}
\end{document}